\documentclass[preprints,article,accept,moreauthors,pdftex]{mdpi} 

\setitemize{parsep=6pt,itemsep=0pt,leftmargin=*,labelsep=5.5mm}
\setenumerate{parsep=6pt,itemsep=0pt,leftmargin=*,labelsep=5.5mm}
\setlist[description]{itemsep=0mm}

\firstpage{1} 
\pubvolume{xx}
\issuenum{1}
\articlenumber{5}
\pubyear{2020}
\copyrightyear{2020}
\history{Received: date; Accepted: date; Published: date}

\pdfoutput=1

\newcommand{\bra}[1]{\left\langle#1\right|}
\newcommand{\ket}[1]{\left|#1\right\rangle}

\newcommand{\ketbra}[2]{\left|{#1}\right\rangle\left\langle{#2}\right|}

\usepackage{bbm} 

\Title{Distance between Bound Entangled States from Unextendible Product Bases and Separable States}

\Author{Marcin Wie\'sniak $^{1,2,}$*, Palash Pandya~$^{1}$, Omer Sakarya~$^{3}$ and Bianka~Woloncewicz $^{1,2}$}

\AuthorNames{Firstname Lastname, Firstname Lastname and Firstname Lastname}

\address{%
$^{1}$ \quad Institute of Theoretical Physics and Astrophysics, Faculty of Mathematics, Physics, and Informatics, University of Gda\'nsk, 80-308 Gda\'nsk, Poland; palashpandya.iiith@gmail.com (P.P.); bibivolo@gmail.com~(B.W.)\\
$^{2}$ \quad International Centre for Theory of Quantum Technologies (ICTQT), University of Gda\'nsk, 80-308 Gda\'nsk, Poland\\
$^{3}$ \quad Institute of Informatics, Faculty of Mathematics, Physics, and Informatics, University of Gda\'nsk, \mbox{80-308 Gda\'nsk, Poland}; osakarya@sigma.ug.edu.pl}

\corres{Correspondence: marcin.wiesniak@ug.edu.pl}

\abstract{We discuss the use of the  Gilbert algorithm to tailor entanglement witnesses for
unextendible product basis bound entangled states (UPB BE states). The method relies on the fact that an optimal entanglement witness is given by a~plane perpendicular to a~line between the reference state, entanglement of which is to be witnessed, and its closest separable state (CSS). The~Gilbert algorithm finds an approximation of CSS. In this article, we investigate if this approximation can be good enough to yield a~valid entanglement witness. 
We compare witnesses found with Gilbert algorithm and those given by Bandyopadhyay–Ghosh–Roychowdhury (BGR) construction. This comparison allows us to learn about the amount of entanglement and we find a~relationship between it and a~feature of the construction of UPBBE states, namely the size of their central tile. We show that in~most studied cases, witnesses found with the Gilbert algorithm in this work are more optimal than ones obtained by Bandyopadhyay, Ghosh, and Roychowdhury. This result implies the increased tolerance to experimental imperfections in a realization of the state.}

\keyword{bound entanglement; entanglement witness; Hilbert–Schmidt measure; optimization algorithms}

\begin{document}


\section{Introduction}
Entanglement is likely the most counter-intuitive feature of quantum mechanics. It allows quantum systems to exhibit correlations, which cannot be reconstructed by any set of prearranged local quantum states. As such, entanglement is seen as a~resource responsible for advantage in various communication tasks (see, e.g., Ref.~\cite{COMCOMP}). However, while it is trivial to describe entanglement for pure states~\cite{ENTPURE}, it is one of the most important open questions of quantum information theory to determine whether a~given mixed state is entangled. This problem is additionally complicated by, e.g., existence of bound entanglement~\cite{BOUND}, which cannot be transformed in the local actions and classical communication (LOCC) regime into an ensemble of pure maximally entangled states. Hence, these states have limited applications in quantum communication. 

For bipartite states with distillable entanglement, we again have a~straightforward tool to detect their nonclassicality: they become non-positive under partial transposition~\cite{PPT}. Thus, the problem of certifying entanglement for the bipartite case is reduced to the case of bound entangled (BE) states~\cite{BOUND}. In theory, any form of entanglement can be certified with a~proper positive, but not completely positive (PNCP) map~\cite{MAPS}. The Jamio\l kowski–Choi isomorphism~\cite{CHOI,JAMIOL} links each such map to a~linear operator called an entanglement witness~\cite{WITNESS}. A reference state, entanglement of which we want to confirm, should give the expectation value of the witness inaccessible with only product states.

Naturally, any entangled state has its own class of witnesses that detect its nonclassical correlations. This follows directly from the fact that it is represented by a~point in some distance from a~{convex} set of separable states---there are infinitely many states separating the two, but for an individual state they cannot define the whole boundary of the set. The witness that detects entanglement in the largest volume of the state set is optimal. Since neither the set of all the states nor of the separable ones is a~polytope, an optimal entanglement witness is related to a~hyperplane tangential to a~line connecting the reference state and CSS. Contrary to a~common belief, an optimal entanglement witness does not correspond to robustness against any particular form of noise.

The plethora~of entanglement witnesses, or equivalently, PNCP maps, is responsible for the complexity of the problem of entanglement detection. Not only do we need to optimize the  expectation value of the witness over all separable states, but also the witness itself needs to be a~subject of optimization, as we have no a~priori knowledge about its form. The {\em max-max} nature of the problem is the main difficulty in tackling the question, even just numerically.

In Ref.~\cite{BGR}, Bandyopadhyay, Ghosh, and Roychowdhury proposed a~construction of entanglement witnesses for UPB BE states. They have a~strikingly simple structure comprised of the combination of a~projection onto the support of the state and the unit operator with a~weight equal to the maximal mean value of the UPB BE state in a~separable state. Unfortunately, till date these witnesses have not been studied extensively, for example, in terms of optimality. From this point onwards, this family of entanglement witnesses will be referred to as {BGR witnesses}.

In our recent article~\cite{PALASH} we have employed the Gilbert algorithm~\cite{GILBERT} to find approximation of a~separable state closest (with respect to the Hilbert–Schmidt measure, $D(\rho,\sigma)=\sqrt{\text{tr}(\rho-\sigma)^2}\,$) to a~reference state. We argued that basing on decay of the distance (and other properties) we can classify a~state as strongly entangled or practically separable. Here, we elaborate on this approach by asking if an implementation of the Gilbert algorithm can yield a~separable state close enough to a~reference state that it gives a~state-tailored entanglement witness. As argued in Ref.~\cite{BDHK}, the optimal entanglement witness for state $\rho_0$ will be given by a~hyperplane perpendicular to a~vector in the state space $\rho_{\text{CSS}}-\rho_0$, where $\rho_{\text{CSS}}$ is the closest separable state with respect to the Hilbert–Schmidt measure. The algorithm cannot reach $\rho_{\text{CSS}}$, but instead it gives an approximation $\rho_1$, which introduces a~small tilt in the plane of the witness. Also, as subsequent corrections to the found state become exponentially small, it is not feasible to reach arbitrarily high precision. Thus, the question is: can this tilt be small enough, so that the hyperplane still defines a~valid entanglement witness for $\rho_0$. 
 
Computationally, our approach is substantially  easier than solving a~{\em max-max} problem. The~latter requires that the two optimizations be conducted in turns. It may happen that a~step taken in optimizing a~witness actually degrades it  which can be only verified by optimizing over separable states. On the other hand, the Gilbert algorithm guarantees convergence to the true solution, so it can be conducted independently. 

To test this hypothesis, we chose  BE states from unextendible product bases~\cite{UPBBE,UPBBE1}. On~one hand, they have a~simple construction, which can be straight-forwardly generalized to a~large number of cases. On the other---their bound entanglement is nontrivial to be certified. Their additional advantage is that their CSS can be found mixing only strictly real states, which improves the efficiency. Also, an entanglement witness can be constructed analytically for them~\cite{BGR}.

In the next section we review the construction of BE states. In Section~\ref{sec3} we briefly discuss the concept of an entanglement witness. Section~\ref{sec4} describes the Gilbert algorithm. The numerical results are discussed in Section~\ref{sec5}, followed by conclusions.

\section{Bound Entangled States from Unextendible Product Bases} \label{sec2}
\begin{figure}[H]

\centering
\includegraphics[width=6cm]{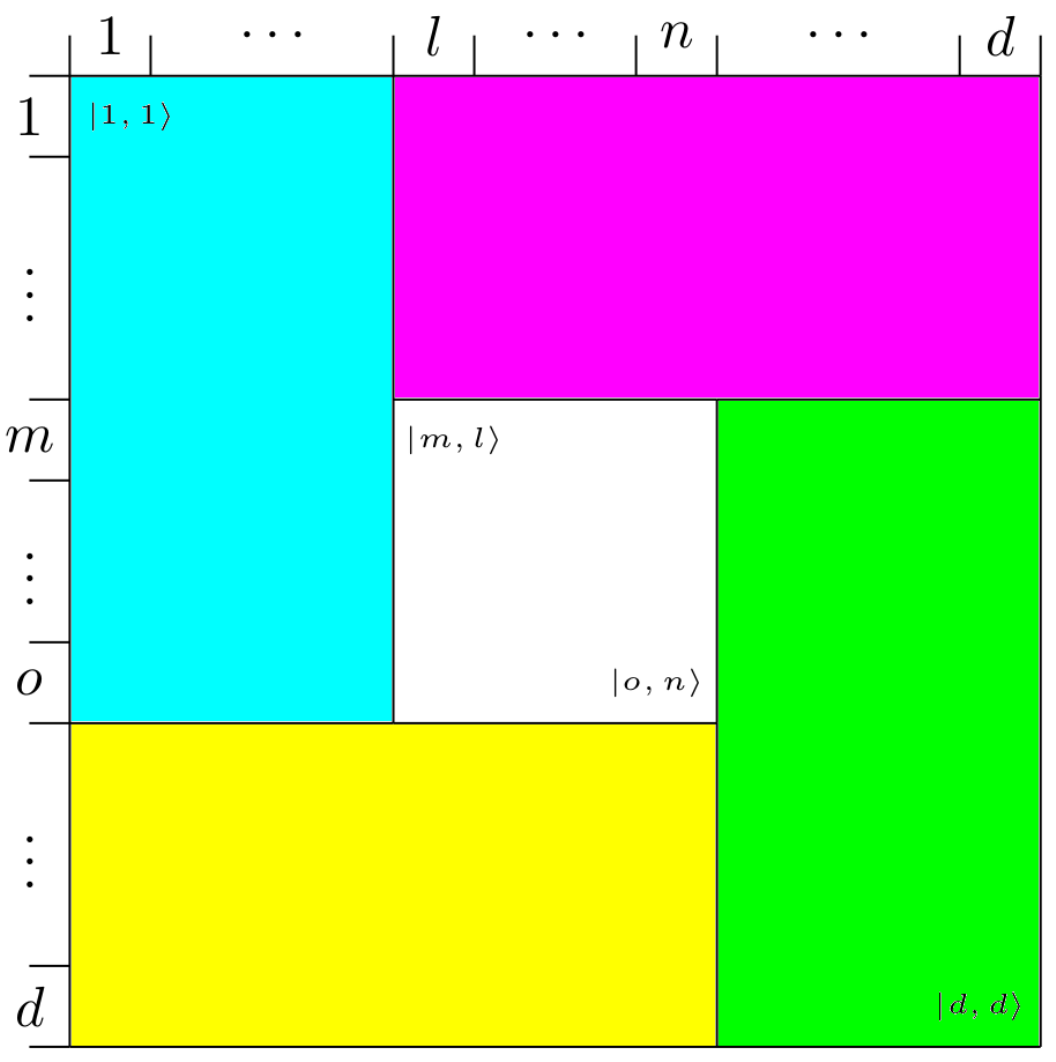}
\caption{Visualization of the structure of bound entangled states from unextendible product bases considered in this article. The size of the central tile is $(n-l+1)(m-o+1)$.} 
\label{fig-UPB-tiles}
\end{figure}

We continue with a~brief {introduction} of bound entangled states from unextendible product bases (UPB BE states). It is one of most fundamental methods of generating bound entangled states, first presented in Refs.~\cite{UPBBE,UPBBE1}. The idea~is as follows. Given a~two- (or more) party Hilbert space we choose a~subspace, which can be represented by a~factorizable set of projectors from the computational basis. Subsequently, we remove from this subspace the projector on the equal superposition of all computational basis states belonging to that region. The remaining is defined as a~tile. For example, in Hilbert space $H=\text{span}(\ket{1},\ket{2},\ket{3})^{\otimes 2}$, a~possible tile would be $\sum_{i,j=1,2}\ketbra{i,j}{i,j}-\frac{1}{2}\sum_{i,j,k,l=1}^2\ket{i,j}\bra{k,l}$. Figure \ref{fig-UPB-tiles} depicts such a construction in $d$-dimensions.  The next step is to find a~covering of the entire Hilbert space such that there is no subset of regions related to tiles, which can be merged into a~region, which is, again, a~factorizable region. Once such a~covering is found, we sum subspaces of all the tiles, and add a~projector on the uniform superposition of all states from the computational basis, $\Pi_{sym}=\frac{1}{d_1d_2}\sum_{i,k=1}^{d_1}\sum_{j.,l=1}^{d_2}\ketbra{i,j}{k,l}$ (sometimes called a~stopper state). As we have previously removed the symmetric states from each tile, they are orthogonal to this state. The bound entangled state is then the normalized projector onto the complement of the sum. The state is PPT, which can be argued from its construction, but it is entangled, since its kernel does not contain any product states. There is no product state orthogonal to the sum of all tiles and the symmetric state anymore. 
	
Let us denote a~projection on the subspace related to a~$k$-th tile as $\Pi^{(k)}$, and onto the symmetrized state of that subspace as $\Pi_{sym}^{(k)}$. Then, any UPB BE state is given by:
\begin{equation}
\rho=\frac{1}{K}(\mathbbm{1}-\Pi_{sym}-\sum_{k=1}^K(\Pi^{(k)}-\Pi^{(k)}_{sym})).
\end{equation}

The rank of the state is hence $K-1$, where $K$ is the number of tiles necessary to cover the whole Hilbert space. For two parties, the minimum number of tiles is 5, while for three parties, 9 tiles are~necessary. 

Possible low rank of the UPB bound entangled states makes them attractive from the experimental point of view. Granted access to a~source of an arbitrary pure state, a~resource readily available with modern high-intensity sources of multidimensional entanglement, local filters and operations, one can relatively easily realize mixtures. One of the greatest challenges in such realizations is high sensitivity of the bound entanglement. For example, any deviation of weights of the eigenstates from the support of the state will transform it into an NPT (negative partial transpose) state. This aspect will play a~role when the construction of the source cannot guarantee equal mixing. A  possible solution is to add white noise, but PPT entanglement tends to be rather fragile.

Another approach is to {\em emulate} the state. Instead of looking at its eigenvectors, our aim is to reconstruct the probability distributions that it  produces. This can be done by taking a~proper combination of product states, possibly with negative coefficients. In case of UPB BE states we can sum up probabilities given by all the states $\Pi^{(k)}_{sym}$ and measure relevant probabilities  $\Pi_{sym}$ to be deducted. Naturally, to establish probabilities we need to measure the projections of complementary states. This can be done after the measurements, or in the following procedure. The observers are informed every time $\Pi_{sym}$ is produced and a~projection is successful, and they increase the denominator in the relative frequency, but decrease the numerator. In some cases, this might lead to ``negative probabilities'', which then should be taken to be zero, or the best fit to a~physical density operator.

Naturally, such an emulated state cannot bring any advantage in communication tasks or distributed computation. In particular, the emulation cost is broadcasting a~bit. However, emulation can be extremely useful for testing and perfecting techniques for 	certifying entanglement. 
As it is complicated to build
a~source of an  entangled state, it is experimentally convenient to first master the accompanying techniques to be sure that the desired goal can be achieved.

\section{Entanglement Witnesses}\label{sec3}

Let us start with a~brief recall of the concept of an entanglement witness. Most generally, it would be a~Hermitian operator with a~certain range of eigenvalues accessible only in entangled states. In that sense, a~Hamiltonian of an antiferromagnetic spin chain of any length is an entanglement witness~\cite{Heisenberg}. By a~common convention, the lowest mean value attainable of a~witness with separable states is 0, and some entangled states may lead to negative mean values. We note, however, that this convention may not always be practical. For any entanglement witness $W$, we can define a~linear functional $\Lambda_W(\cdot)=\text{tr}(\cdot W)$, which divides the space of Hermitian operators with a~hyperplane. On one side of this hyperplane $(\Lambda_W(\cdot)\geq 0)$ lie all separable states, on the other $(\Lambda_W(\cdot)< 0)$---our reference state $\rho_0$  { and its neighborhood}. Since the set of separable states in a~given Hilbert space is convex, there always exist such a~hyperplane for any entangled state. 

Then we can introduce a~class of tight witnesses for which condition $\langle W\rangle_{\text{SEP}}=0$, i.e., the mean value of the witness in the set of separable states is equal to 0, is  achieved for one or more separable states. Within this class, for any entangled $\rho_0$ we can define optimal witness $W_0$, for which $\langle W_0\rangle_{\text{SEP}}=0$ is attained for the closest separable state to 
$\rho_0$ in the sense of Hilbert–Schmidt norm. As a~result, the distance between $\rho_0$ and the hyperplane $\Lambda_{W_0}(\cdot)=0$ is the largest among all witnesses detecting entanglement of $\rho_0$. However, contrary to  common belief, this  does not correspond to the highest robustness against the noise of any universal form.  Our goal is to confirm the form of such witnesses for UPB BE states.

A 5-tile bipartite UPB BE states with a~fixed heliocity can be characterized by dimensions $d_1,d_2$ and coordinates of the central tile $(l,n), (m,o)$ with $1<l\leq n<d_1$ and $1<m\leq o<d_2$. We have studied cases of $d_1=d_2\equiv d=3,4,5,6$. This gives a~single $3\times3$ state, nine $4\times 4$ states, $36$ $5\times 5$ states, and 100 $6\times 6$ states, giving $146$ density matrices in total. In general, in dimensions $d_1\times d_2$ we will have $\frac{1}{4}(d_1^2-3d_1+1)(d_2^2-3d_2+1)$ states.

\section{The Gilbert Algorithm}\label{sec4}
According to Bertlmann–Durstberg–Heismayr–Krammer theorem~\cite{BDHK}, the optimal entanglement witness for state $\rho$ is given by the difference of $\rho$ and the closest separable state with respect to Hilbert–Schmidt measure, $W\propto \rho-\rho_{CSS}$. Finding  $\rho_{CSS}$ is generally a~complex task, but a~very good approximation can be found with the Gilbert algorithm~\cite{GILBERT}. The application of the algorithm to the classification of states as entangled was discussed in Ref.~\cite{PALASH}. In short, the algorithm is as follows:

{{Input}: test state $\rho_0$, arbitrary separable state $\rho_1$,}

{{Output}: approximation of a~separable state closest to $\rho_0$} 

\begin{enumerate}
\item{Choose at random a~pure product state $\rho_2$.}
\item{Maximize $\text{tr}(\rho_0-\rho_1)(\rho_2-\rho_1)$, or go to step 1 if $\text{tr}(\rho_0-\rho_1)(\rho_2-\rho_1)\leq 0$.}
\item{Find the point $\rho'_1$, which lies on line $\rho_1d-\rho_2$, and minimizes the Hilbert–Schmidt distance $D(\sigma,\rho)=\sqrt{\text{tr}(\rho-\sigma)^2}$.}
\item{Update $\rho_1\leftarrow \rho_1'$.}
\item{Go to point 1 until a~given HALT condition is met.}
\end{enumerate}

Let us note that local states constituting $\rho_2$ are drawn with the Haar measure, according to an algorithm by \.Zyczkowski and Sommers~\cite{ZS}. The Hilbert–Schmidt measure has been chosen as the algorithm then requires only	solving a~quadratic equation, although it was shown that it is not monotonous under LOCC and thus is not a true entanglement measure~\cite{WT,OZAWA}. A typical choice of the HALT condition is simply a~time constrain, number of trial states ($\rho_2$), or number of corrections. 

The Gilbert algorithm can provide an accurate approximation of the closest separable state, but, in fact, we cannot reach that state. For this reason, we only partially succeeded in finding almost optimal entanglement witness. The procedure is the following.  We run the algorithm, which halts at some number of corrections of $\rho_1$. We then optimize the mean value of $\rho_0-\rho_1$ for separable states. When this mean value is lower than $\text{tr}(\rho_0-\rho_1)\rho_0$, the witness is found. For  We have found a~witness for the $3\times 3$ state after 25,100 corrections (time constrain), for $4\times 4$ and $5\times 5$ states we run the algorithm for up to 4000 corrections, while for $6\times 6$ states we have conducted only 3500 corrections (number of corrections). 

The first criterion of a~state being entangled or separable is to investigate the limit of distance between $\rho$ and the closest separable state found at the given step. This distance was registered every 50~corrections. As argued in Ref.~\cite{PALASH} there is a~strong linear dependence between number of corrections $c$ and $|\log(\text{tr}(\rho_1-\rho_0)^2-a)|^b$ for some values of $a$ and $b$. We hence maximize the linear regression coefficient between these two quantities. The found value of $b$ is irrelevant, while $a$ is an approximation of the limit. The state can be considered entangled if $a$ is above the precision of the algorithm, estimated to be less than $10^{-5}$.

\section{Numerical Results}\label{sec5}

All 146 UPB BE states have been shown to have the estimated distance to the set of separable sates between 0.09 and 0.06, therefore all of them can be classified as entangled as they indeed are from the construction. On the other hand, it is difficult to argue that any one of them is particularly strongly~nonclassical.

In the next step, we will attempt to construct an entanglement witness for each state. In~a~perfect case of the algorithm actually reaching the closest separable states $\rho_{CSS}$, the witness would be simply proportional to $\rho_{CSS}-\rho_{0}-\text{tr}\rho_{CSS}(\rho_{CSS}-\rho_{0}) $. However, as mentioned above, we reach only an~approximation $\rho_1$ of $\rho_{CSS}$, which causes the hyperplane of the witness to be tilted. We~thus need to conduct an optimization over the set of separable states, $|A\rangle\otimes |B\rangle$, so that $\lambda=\max_{|A\rangle,|B\rangle}\langle A|\langle B|\rho_1-\rho_0|A\rangle|B\rangle$. Then the witness reads:
\begin{equation}
W(\rho_0)=\rho_1-\rho_0-\mathbbm{1}\lambda.
\end{equation}

It turns out that in almost all cases $\rho_1$ yielded a~valid entanglement witness. When transformed to a~traceless form, $\rho_1-\rho_0$, each of these witnesses has a~similar structure: there are four negative eigenvalues of roughly the same magnitude and the corresponding eigenspace is approximately the support of $\rho_0$, and hence $W(\rho_0)$s can be compared with witnesses  in Ref.~\cite{BGR}. A significant difference is that positive eigenvalues are not uniform.

It now remains to compare the strength of the witnesses. Because the witnesses found with the Gilbert algorithm are approximations of the witnesses $W'(\rho_0)$, constructed accordingly to Ref.~\cite{BGR}, we~would focus on the latter. The simplest quantizer of nonclassicality of UPB BE states would be the Hilbert–Schmidt distance of the state to the hyperplane defining the witness. To establish this quantity, we need vector $\vec{M}$ tangent to the hyperplane defining the witness, which in our case is the traceless part of the witness operator, $\vec{M}=W'(\rho)-\mathbbm{1}\text{tr}(W'(\rho_0))$. The trace part is removed from the witness operator to ensure that the corresponding vector starts at the origin of the coordinates system. We~will also need state $\rho'$ saturating $\text{tr} \rho'W'(\rho_0)=0$. This state is found in the process of constructing the witness. Let $\vec{R}=\rho_0-\rho'$. Then, our distance to separable states is the length of the projection of $\vec{R}$ onto unit vector $\vec{M}/\sqrt{\vec{M}\cdot\vec{M}}$ is given by:
\begin{equation}
\label{projection}
D(\rho_0,\text{Sep})=\frac{|\vec{R}\vec{M}|}{\sqrt{|\vec{M}|^2}}=\frac{\left|\text{tr} W'(\rho_0)(\rho_0-\rho')\right|}{\sqrt{\text{tr}W'^2(\rho_0)}},
\end{equation}
where Sep denotes the set of separable states.

\begin{figure}[H]
\centering
\includegraphics[width=6cm]{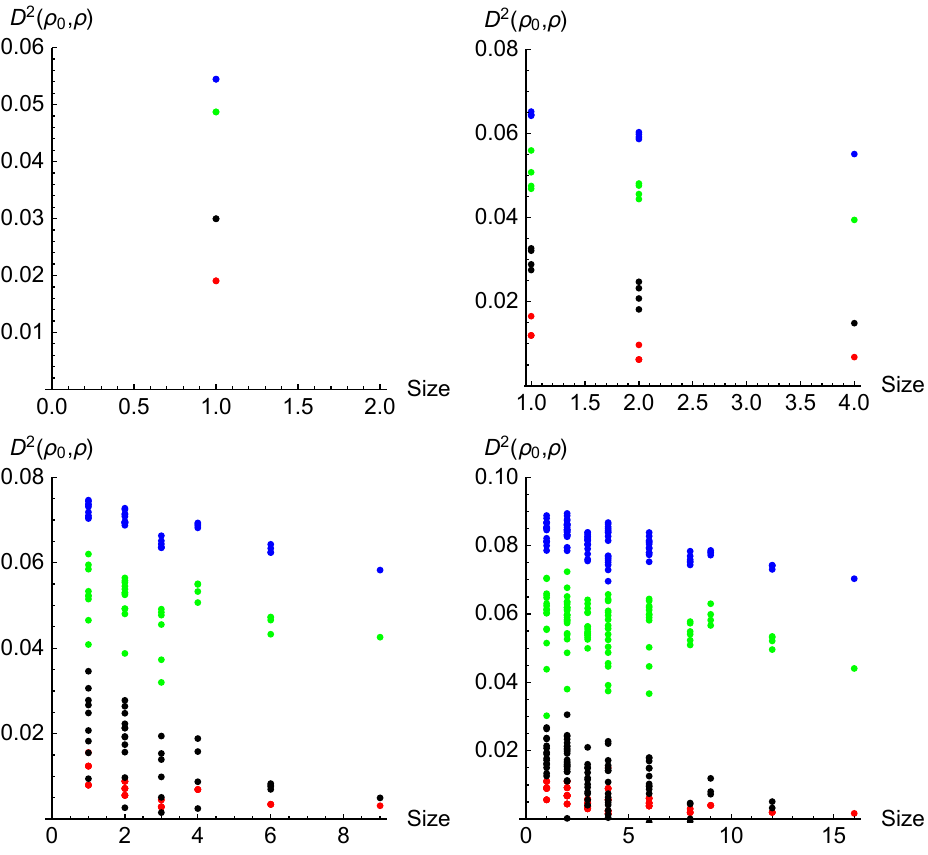}
\caption{The comparison between last distance between a~UPB BE state found by the Gilbert algorithm after 25,100 corrections ($3\times 3$, top left), 4000 corrections ($4\times 4$, top right, and $5\times 5$, bottom left), and 3500 corrections ($6\times 6$, bottom right). Red points correspond to distances from Equation~(\ref{projection}) with  BGR witnesses, the black points show the distance to the hyperplanes of the witnesses found by the Gilbert algorithm. The last distance found by the algorithm is marked with blue points and the extrapolation of its decay is represented with green points. The data~are segregated by the size of the central tile.}\label{fig2}
\end{figure}

Figure~\ref{fig2} presents comparison of four estimates of the distance between five-tile UPB BE states and separable states. Blue points correspond to the smallest distance found by the Gilbert algorithm after the mentioned number of corrections, green points show the estimate of the distance from the linear regression, red represent  the distance computed from Equation~(\ref{projection}) for the BGR witnesses, while black points illustrate the distance given by the witnesses found by our algorithm (again, computed from Equation~(\ref{projection})). The first quantity is an  upper bound of the actual distance, the second is just an~estimate, while the distances computed from the witnesses are lower bound.

It should be noted that in all four plots, the red points are significantly below all other groups. This clearly indicates that we have found more optimal witnesses, than those given in Ref.~\cite{BGR}. Another observation, for every state, the green point is above the black one, i.e., the decay estimation is between the upper and the lower bound. However, our algorithm did not yield a~valid witness for 13 states in $6\times 6$ Hilbert space. 
Also, for 6 states in  $6\times 6$ and  4 states in $5\times 5$ Hilbert spaces, the witness yielded by our algorithm was weaker than the BGR witnesses. 
Still, our algorithm is successful for most tested states. Furthermore, our  results show that BGR construction does not lead to an optimal witness for UPB BE states, as we find more optimal ones.
  
\section{Conclusions}\label{sec6}
In this contribution, we have studied quantifiable entanglement of bound entangled states derived from unextendible product bases in dimension  $d\times d$, where $d=3,4,5,6$. As a~prime tool, we have used the Gilbert algorithm to find an approximation of a~separable state closest to the given entangled state. Knowledge of this approximation leads to a~close-to-optimal entanglement witness. Our method succeeded to yield entanglement witnesses in 133 of the 146 studied cases. Moreover, witnesses found by our algorithm were in 123 cases more optimal than those proposed by Bandyopadhyay, Ghosh, and Roychowdhury~\cite{BGR}.  Both the decay of the distance with successive corrections, and its extrapolation by a~linear fit indicates a~presence of entanglement.  
Also, these results show that construction of BGR  witnesses is not optimal. There exist convex combinations of $\rho_0$ and $\rho_{CSS}$ or $\rho_1$, which are not recognized as entangled by BGR witnesses, whereas our algorithm  reveals quantum correlations in such~cases. 

We have calculated the distance between PPT states and the sets of separable states. UPB BE states turned out to be relatively close to the set of separable states, with all distances below 0.1. While no clear relationship between the structure of the state and the distance was established, one correlation appears, i.e., states with smaller central tiles lay closer to separable states.

Our results therefore have multifold scientific aspects. First, we have partially found the order of degree of entanglement of UPB BE states, by recognizing that in the given dimension, those farthest from the set of separable states have a~smaller central tile. Second, the witnesses yielded by the algorithm are (in most cases) more optimal than the BGR construction. As a~consequence, our witnesses allow for larger imperfections in an experimental realization. This could be relevant in quantum randomness amplification and cryptographic key distribution protocols. 

\vspace{6pt}


\funding{This research was funded by NCN Grants 2017/26/E/ST2/01008, 2015/19/B/ST2/01999 (MW),  2014/14/E/ST2/00020 (PP), and the ICTQT IRAP project  of FNP, financed by structural funds of EU.}

\acknowledgments{We thank Remigiusz Augusiak for useful insight on this publication.}


\reftitle{References}

\end{document}